\documentclass[conference]{IEEEtran}

\usepackage[utf8]{inputenc}
\usepackage[pdftex]{graphicx}
\usepackage{amsmath,amssymb}
\usepackage{physics}
\usepackage{array}
\usepackage[caption=false]{subfig}
\usepackage{stfloats}
\usepackage{url}
\usepackage{xcolor}
\usepackage{verbatimbox}
\usepackage{xstring}

\newcommand{\argmax}{\mathop{\text{argmax}}}

\newcommand{\cpc}{\text{CP}}

\newcommand{\note}[2]{
    \IfEqCase{#1}{
        {max}{\textcolor{cyan}{#2}}
        {ben}{\textcolor{orange}{#2}}
        {david}{\textcolor{violet}{#2}}
    }[\PackageError{note}{Undefined option to note: #1}{}]%
}

\def\eq#1{Eq.~\ref{eq:#1}}
\def\fig#1{Fig.~\ref{fig:#1}}
\def\sec#1{Sec.~\ref{sec:#1}}

\begin{document}
    \title{Depth versus Breadth in Convolutional Polar Codes }

    \author{\IEEEauthorblockN{Maxime Tremblay${}^{1*}$, Benjamin Bourassa${}^1$ and David Poulin${}^{1,2}$}
    \IEEEauthorblockA{${}^1$Département de physique \& Institut quantique, Université de Sherbrooke,
     Sherbrooke, Québec, Canada J1K 2R1\\
     ${}^2$Canadian Institute for Advanced Research, Toronto, Ontario, Canada M5G 1Z8 \\
    ${}^*$maxime.tremblay9@usherbrooke.ca}}

    \maketitle

    \begin{abstract}
        Polar codes were introduced in 2009 by Arikan as the first efficient encoding and decoding scheme that is capacity achieving for symmetric binary-input memoryless channels. Recently, this code family was extended by replacing the block-structured polarization step of polar codes by a convolutional structure. This article presents a numerical exploration of this so-called \textit{convolutional polar codes} family to find efficient generalizations of polar codes, both in terms of decoding speed and decoding error probability. The main conclusion drawn from our study is that increasing the convolution depth is more efficient than  increasing the polarization kernel's breadth as previously explored.
    \end{abstract}

    \IEEEpeerreviewmaketitle

    \section{Introduction}
    \label{sec_intro}
        Polar codes build on channels polarization to efficiently achieve the capacity of symmetric channels (refer to \cite{arikan_channel_2009,arikan_rate_2009,sasoglu_polarization_2009, korada_polar_2010} for detailed presentations). Channel polarization is a method that takes two independent binary-input discrete memoryless channels $W(y|x)$ to a \textit{bad} channel and a \textit{good} channel, given by
        \begin{align}
            W(y_1^2|u_1) &= \sum_{u_2 \in \qty{0,1}}W(y_2 | u_2) W(y_1 | u_1 \oplus u_2), 
            \label{eq:bad_channel} \\
            W(y_1^2, u_1 |u_2) &= W(y_2 | u_2) W(y_1 | u_1 \oplus u_2)
            \label{eq:good_channel}
        \end{align}
        respectively, where $x_a^b = (x_a, x_{a+1} \ldots x_b)^\top$. These channels are obtained by combining two copies of  $W(y|x)$ with a CNOT gate $(u_1,u_2)\rightarrow (u_1\oplus u_2, u_2)$ and then decoding successively bits $u_1$ and $u_2$. That is, output bit $u_1$ is decoded first assuming that $u_2$ is erased. Then bit $u_2$ is decoded taking into account the previously decoded value of $u_1$.    

\begin{figure}[!t]
            \centering
            \subfloat[]{
                \centering
                \includegraphics[width=1.25in]{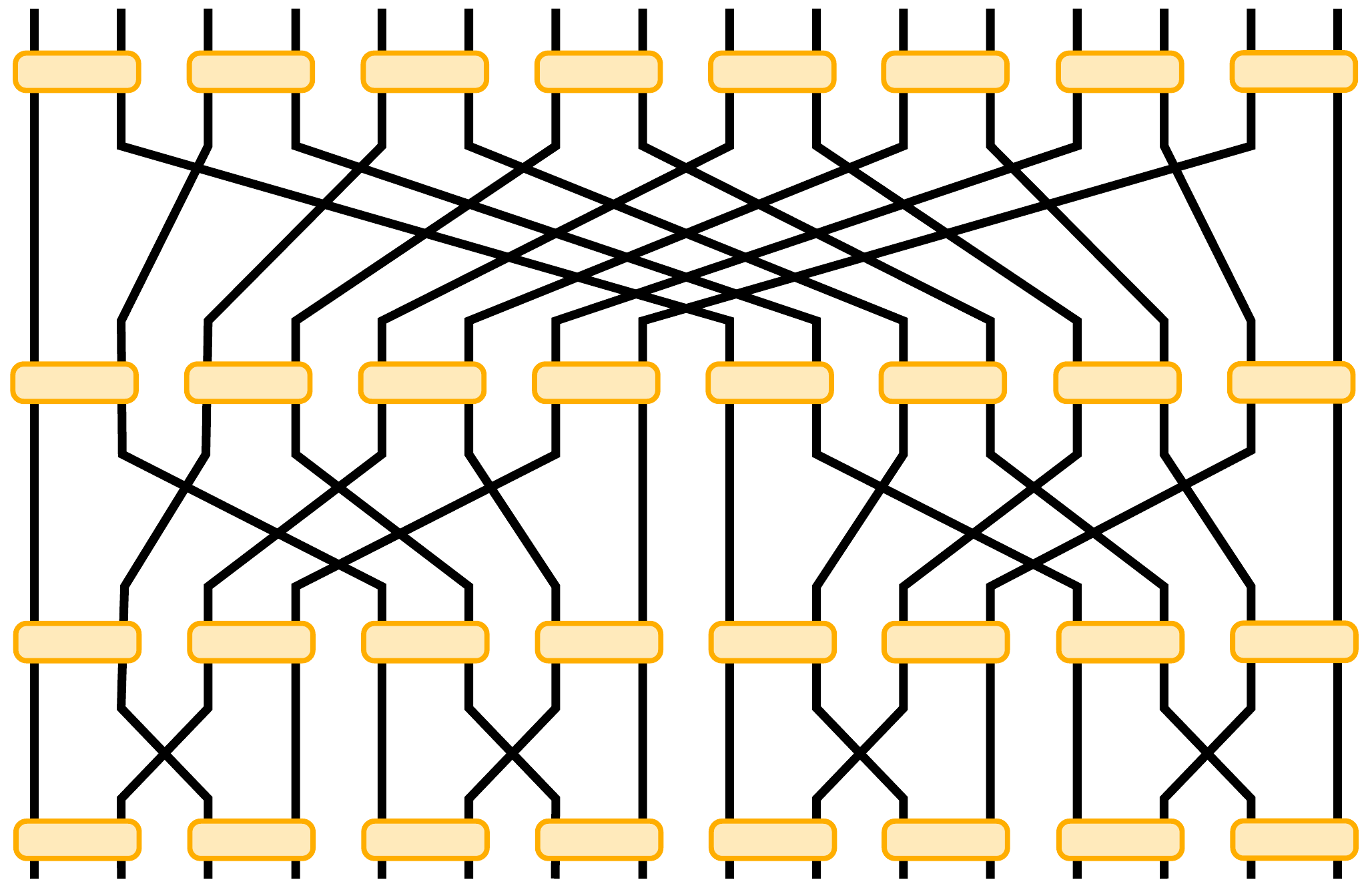}
                \label{fig:subfig_circ_2_1}}
            \quad
            \subfloat[]{
                \centering
                \includegraphics[width=1.83in]{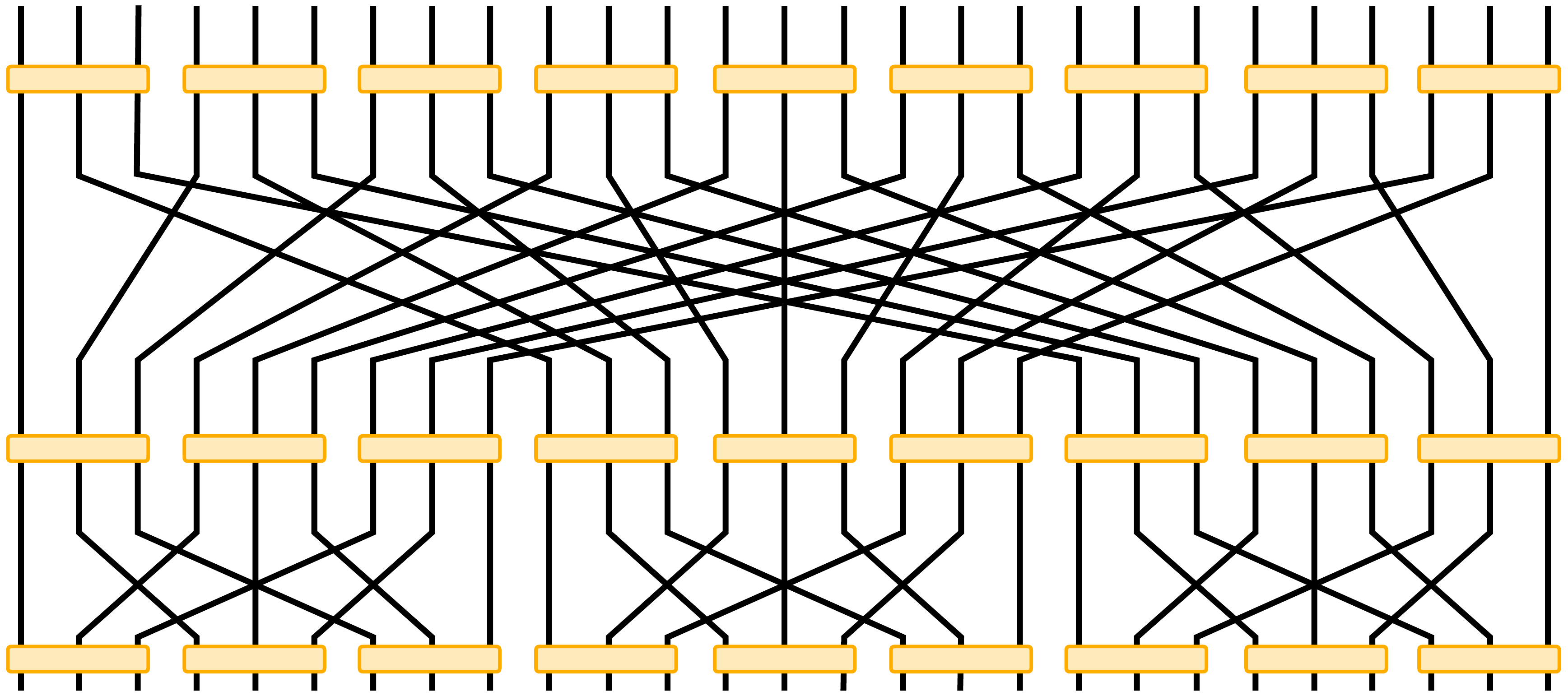}
                \label{fig:subfig_circ_3_1}}
            \quad \\
            \subfloat[]{
                \centering
                \includegraphics[width=1.25in]{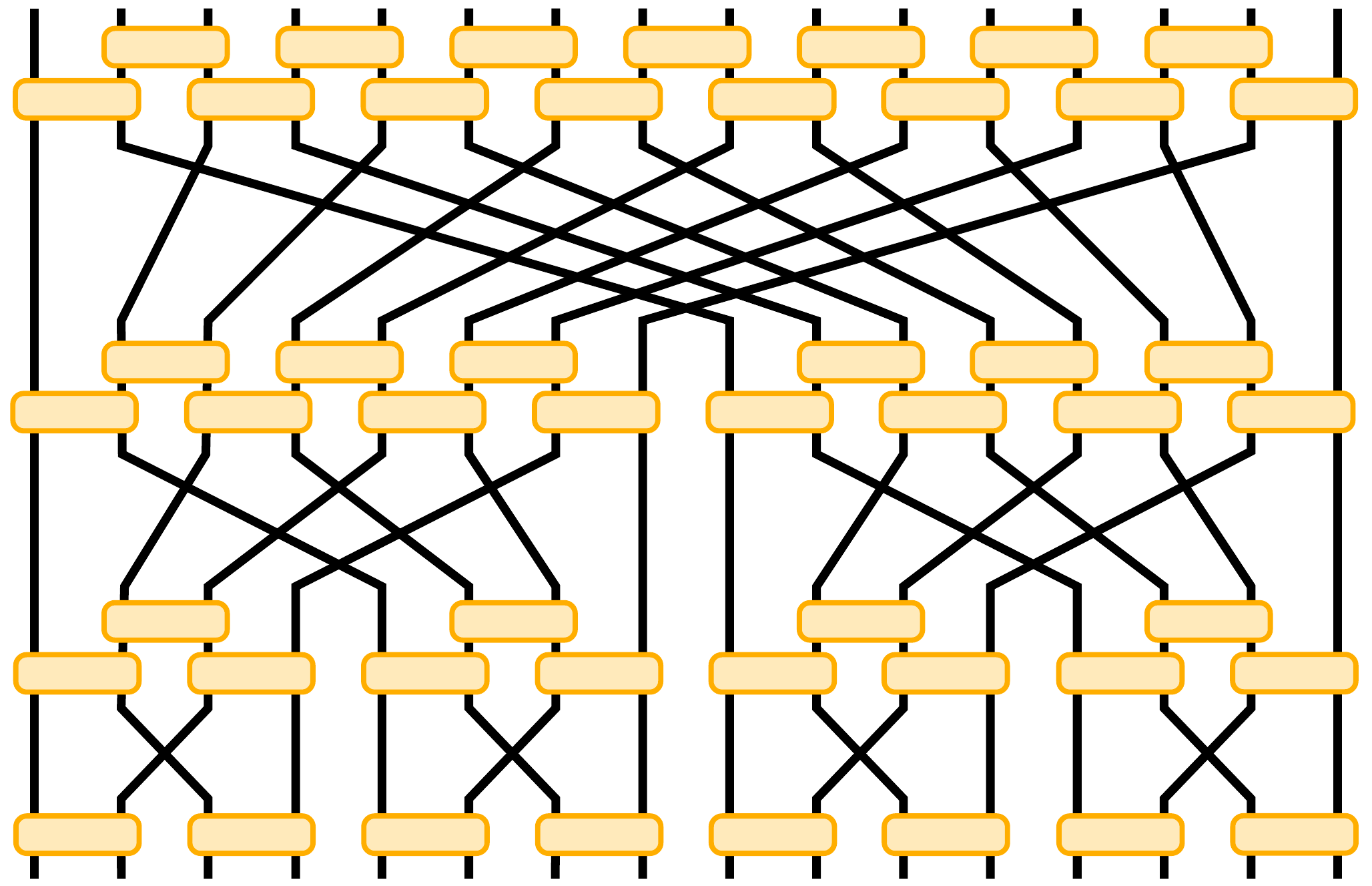}
                \label{fig:subfig_circ_2_2}}
            \quad
            \subfloat[]{
                \centering
                \includegraphics[width=1.83in]{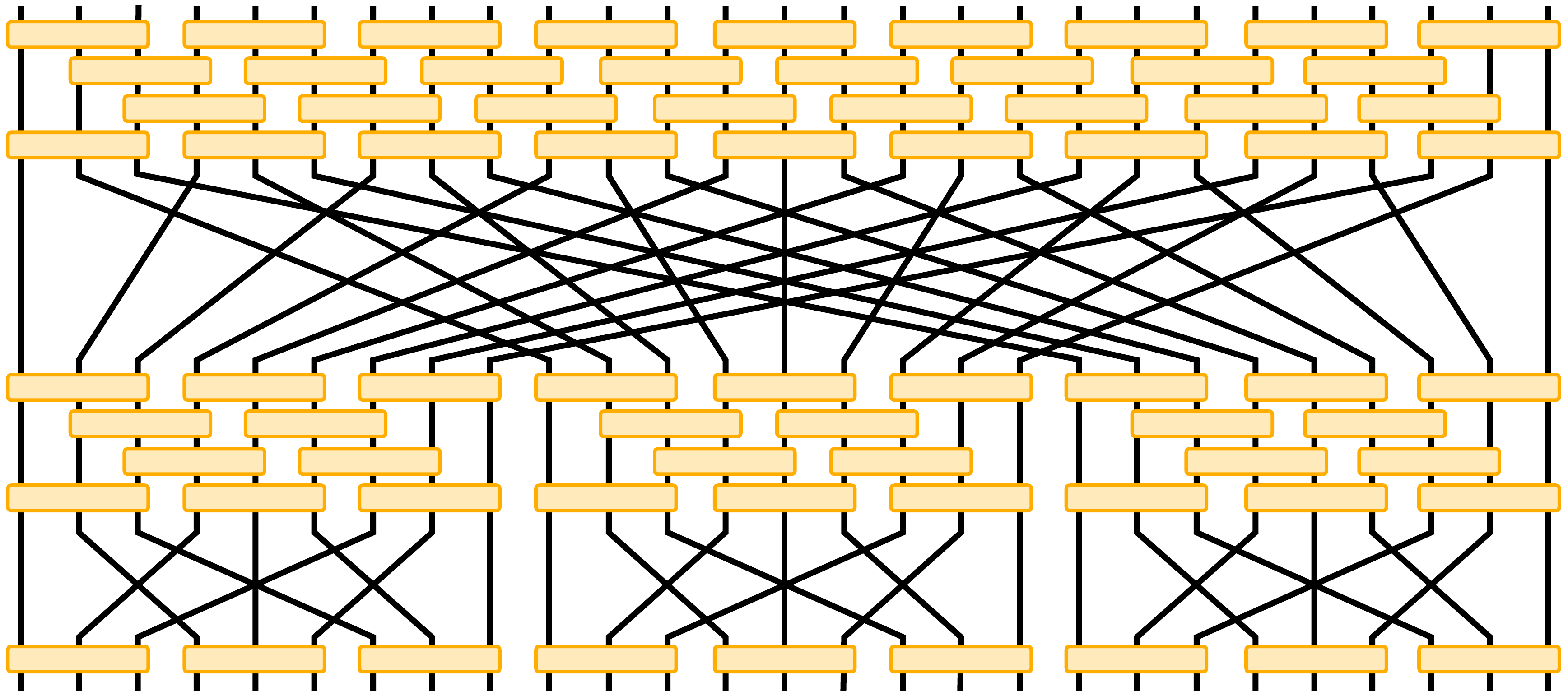}
                \label{fig:subfig_circ_3_4}}
            \caption{Examples of regular (depth=1) and convolutional (depth$>1$) polar code circuits. The parameters (breadth, depth, polarization steps) are  \textbf{(a)} (2,1,4), \textbf{(b)} (3,1,3), \textbf{(c)} (2,2,4) and \textbf{(d)} (3,3,3).}
            \label{fig:fig_circuits}
        \end{figure}

        Polar codes are obtained  by recursing this process to obtain $2^l$ different channels from the polarization of $2^{l-1}$ pair of channels (\fig{subfig_circ_2_1}). As the number of polarization steps $l$ goes to infinity, the fraction of channels for which the error probability approaches 0 tends to $I(W)$ and the fraction of channels for which the error probability approaches 1 tends to $1 - I(W)$, where $I(W)$ is the mutual information of the channel with uniform distribution of the inputs \cite{arikan_channel_2009}. Thus, polar codes are capacity achieving for those channels.
        
        The above construction can be generalized by replacing the CNOT transformation by a different polarization kernel \cite{korada_polar_2010-1}. See \sec{kernels} for details. The kernel can generally take as input more than two copies of the channel $W(y|x)$ and the {\em breadth} $b$ of a kernel is define as the number of channels it combines. An increasing breadth offers the possibility of a more efficient polarization (i.e. a lower decoding error probability), but has the drawback of an increased decoding complexity.

        Another possible generalization of polar codes is to replace the block-structure polarization procedure by a convolutional structure. See \sec{conv} for details.  Note indeed that each polarization step of a polar code consists of independent application of the polarization kernel on distinct blocks of $b$ bits (pairs of bits in the above example with $b=2$). Recently (\cite{ferris_branching_2014,ferris_convolutional_2017}), this idea was extended to a convolutional structure  (see \fig{subfig_circ_2_2} and \fig{subfig_circ_3_4}), where each polarization step does not factor into a product of independent transformations on disjoint blocks but instead consists of $d$ layers of shifted block transformations. We refer to the number of layers $d$ as the {\em depth} of a code. An increasing depth offers the advantage of faster polarization and the drawback of an increased decoding complexity.

        The focus of the present work is to compare the trade-off between breadth and depth in terms of the speed at which the decoding error rate goes to zero and the decoding complexity. We focus on codes which have practically relevant sizes using Monte Carlo numerical simulations. 
       

    \section{Decoding}

        In this section, the general successive cancellation decoding scheme is define  in terms of tensor networks. This enables a straightforward extension to convolutional polar codes.

        \subsection{Successive cancellation}
        \label{sec:SC}

\begin{figure}[!t]
                \centering
                \subfloat[]{
                    \centering
                    \includegraphics[width=0.95in]{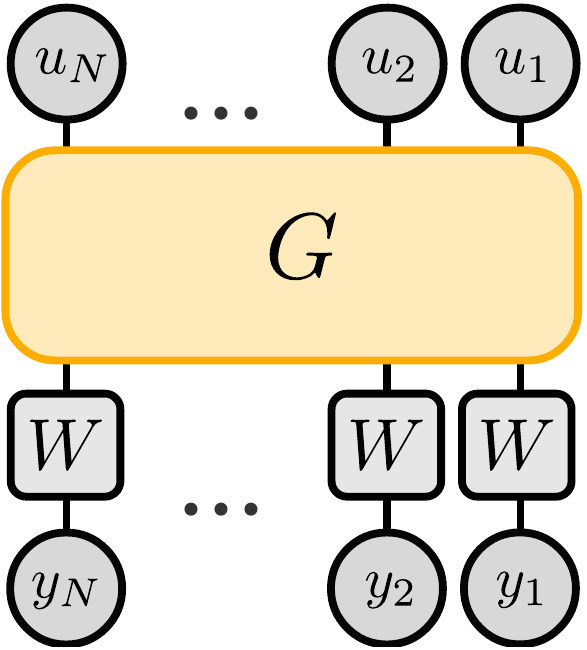}
                    \label{fig:subfig_comp_channel}}
                \qquad
                \subfloat[]{
                    \centering
                    \includegraphics[width=1.69in]{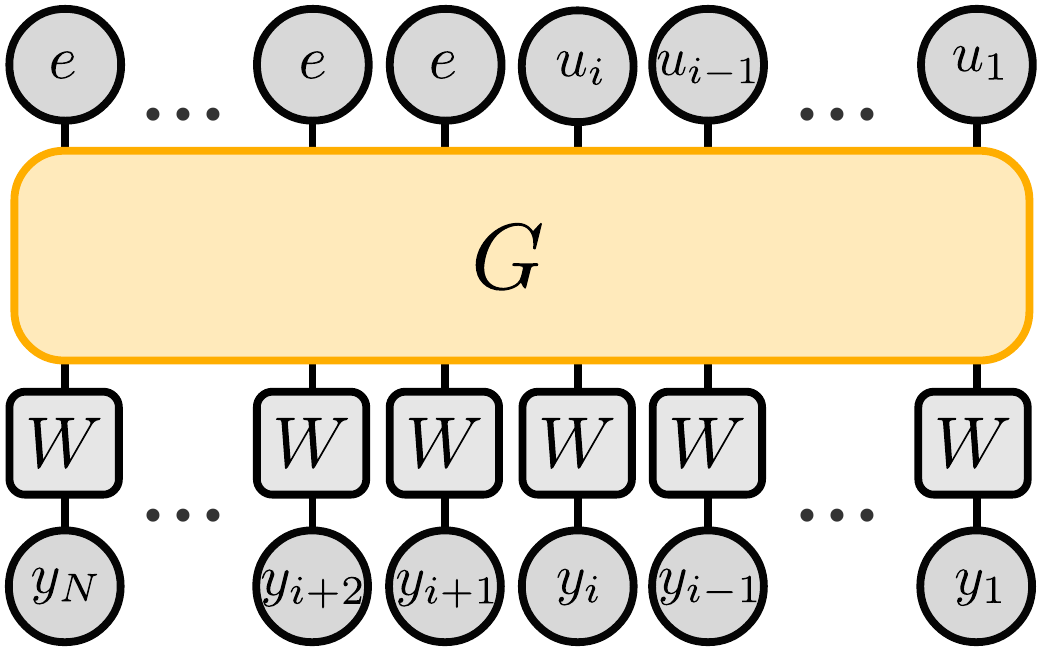}
                    \label{fig:subfig_eff_channel}}
                \caption{Schematic representation of the successive cancellation decoder. \textbf{(a)} A composite channel is obtain from an encoding circuit $G$ and $N$ copies of a channel $W$. Contracting this tensor network for given $y_1^N$ and $u_1^N$ yields \eq{comp_channel}. \textbf{(b)} An effective channel is obtain from the composite channel by summing over all the values of bits $u_{i+1}^N$, graphically represented by the uniform tensor $e = \binom 11$, when decoding bit $u_i$. Contracting this tensor yields \eq{eff_channel} up to a normalization factor.}
                \label{fig:fig_succ_dec}
            \end{figure}

            Define $G$ as the reversible encoding circuit  acting on $N$ input bits and $N$ output bits. $K$ of these input bits take arbitrary values $u_i$ while the $N-K$ others are frozen to the value $u_i=0$. From this input $u_1^N$, the message $x_1^N = G u_1^N$ is transmitted. The channel produces the received message  $y_1^N$,  resulting in a composite channel
            \begin{align}
                W_G(y_1^N| u_1^N) = \prod_{i=1}^N W(y_i | (G u_1^N)_i). \label{eq:comp_channel}
            \end{align}
            This composite channel induces a correlated distribution on the bits $u_i$ and is represented graphically on \fig{subfig_comp_channel}. 

            Successive cancellation decoding converts this composite channel into $N$ different channels given by
            \begin{align}
                W^{(i)}_G(y_1^N, u_{1}^{i-1}| u_i) = \sum_{u_{i+1}, \ldots u_{N}} W_G(y_1^N | u_1^N), \label{eq:eff_channel}
            \end{align}
            for $i = 1,2,\ldots N$. Those channels are obtain by decoding successively symbols $u_1$ through $u_N$ (i.e., from right to left on \fig{fig_succ_dec}) by summing over all the bits that are not yet decoded and fixing the value of all the bits $u_1^{i-1}$. Either to their frozen value, if the corresponding original input bit was frozen, or to their previously decoded value. This effective channel is represented graphically on \fig{subfig_eff_channel}.  
        
            Given $W^{(i)}_G$, $u_i$ is decoded by maximizing the likelihood of the acquired information:
            \begin{align}
                u_i = \argmax_{\tilde u_i\in\qty{0,1}} \, W^{(i)}_G(y_1^N, u_{1}^{i-1}| \tilde u_i). \label{eq_ml_decoder}
            \end{align}
            Applying this procedure for all bits from right to left yield the so-called \textit{successive cancellation decoder}. 
     
            Equation \ref{eq_ml_decoder} can be generalized straightforwardly by decoding not a single bit $u_i$ at the time but instead a $w$-bit sequence $u_i^{i+w-1}$ jointly, collectively viewed as a single symbol from a larger alphabet of size $2^w$. To this effect, the decoding width $w$ is defined as the number of bits that are decoded simultaneously. 

        \subsection{Decoding with tensor networks}
        \label{sec:TN}
        
            Convolutional polar codes were largely inspired by tensor network methods used in quantum many-body physics (see e.g. \cite{orus_practical_2014} and \cite{bridgeman_hand-waving_2017} for an introduction). Akin of the graphical tools used in information theory (Tanner graph, factor graph, etc.), tensor networks were introduced as compact graphical representation of probability distributions (or amplitudes in quantum mechanics) involving a large number of  correlated variables. Moreover, certain computational procedures are more easily cast using these graphical representations. It is the case of the successive cancellation decoding problem described above, where the goal is to compute $W^{(i)}_G(y_1^N, u_{1}^{i-1}| u_i)$ given fixed values of $y_1^N, u_{1}^{i-1}$.
        
            While $G$ is a $\mathbb F_2^N$ linear transformation, it is sometime convenient to view it as a linear transformation on the space of probability over $N$-bit sequences, i.e., the  linear space $\mathbb R^{2^N}$ whose basis vectors are labeled by all possible $N$-bit strings.  On this space,  $G$ acts as a permutation matrix mapping basis vector $u_1^N$ to basis vector $x_1^N = Gu_1^N$. A single bit is represented in the state $0$ by $u=\binom 10$, in the state $1$ by $u=\binom 01$ and a bit string $u_1^N$ is represented by the $2^N$ dimensional vector $u_1^N = u_1\otimes u_2\otimes \ldots \otimes u_N$. A single bit channel is a $2\times 2$ stochastic matrix and a CNOT gate is given by 
            \begin{equation}
                {\rm CNOT} = 
                \left(
                \begin{array}{cccc}
                    1&0&0&0\\
                    0&1&0&0\\
                    0&0&0&1\\
                    0&0&1&0
                \end{array}
                \right),
                \label{eq:CNOT}
            \end{equation}
            because it permutes the inputs $10$ and $11$ while leaving the other inputs $00$ and $01$ unchanged.
     
            In this representation
            \begin{align}
                W^{(i)}_G&(y_1^N, u_{1}^{i-1}| u_i) =    \nonumber
                \\ &\frac{1}{Z}[u_1 \otimes \ldots \otimes u_{i-1} \otimes u_i \otimes e^{\otimes(N-i)}]^T  G W^{\otimes N} y_1^N, \label{eq:Wi}
            \end{align}
            where $e = \binom 11$ and $Z = \sum_{u_i \in \qty{0,1}} W^{(i)}_G(y_1^N, u_{1}^{i-1}| u_i)$ is a normalization factor. Ignoring normalization, this quantity   can be represented graphically as a tensor network (see \fig{subfig_eff_channel}), where each element of the network is a rank-$r$ tensor, i.e., an element of $\mathbb R^{2^r}$. Specifically, a bit $u_i$ is a rank-one tensor, a  channel $W$ is a rank-two tensor, and a two-bit gate is a rank-four tensor (two input bits and two output bits). The CNOT gate is obtained by reshaping \eq{CNOT} into a $(2 \times 2 \times 2 \times 2)$ tensor. 
            
            In this graphical representation, a rank-$r$ tensor $A_{\mu_1,\mu_2,\ldots \mu_r}$ is represented by a degree-$r$ vertex, with one edge associated to each index $\mu_k$. An edge connecting two vertices means that the shared index is summed over
            \begin{equation}
                \includegraphics[width=6.5cm]{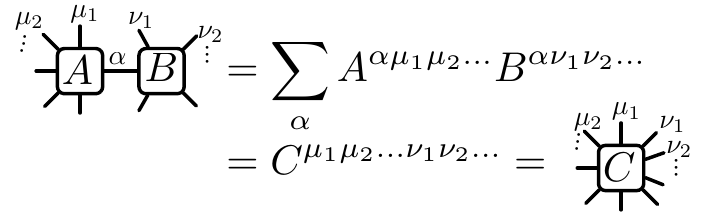},\label{eq:TNC}
            \end{equation}
            generalizing the notion of vector and matrix product to higher rank tensors. Tensors can be assembled into a network where edges represent input-output relations just like in an ordinary logical circuit representation. Evaluating \eq{Wi} then amounts to summing over edge values.
      
\begin{figure}[!t]
                \centering
                \subfloat[]{
                    \centering
                    \includegraphics[width=1.57in]{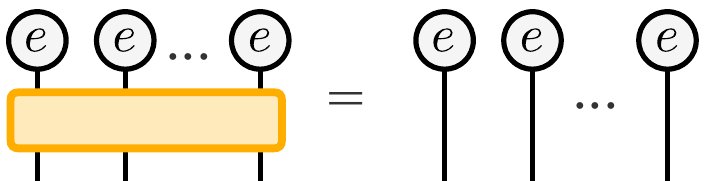}
                    \label{fig:tn_simp_e}}
                \quad
                \subfloat[]{
                    \centering
                    \includegraphics[width=1.57in]{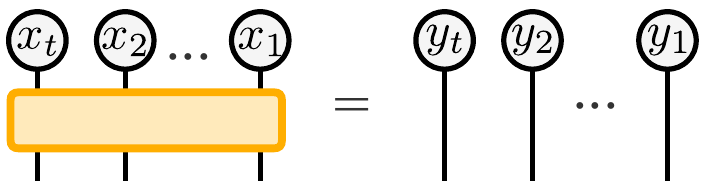}
                    \label{fig:tn_simp_xy}}
                \caption{Circuit identities. \textbf{(a)} Any permutation acting on the uniform distribution return the uniform distribution. \textbf{(b)} Any contraction of a permutation and basis vector $x_1^t$ gives another basis vector $y_1^t$.}
                \label{fig:tn_simp}
            \end{figure}

            This computational task, named {\em tensor contraction},  generally scales exponentially with the tree-width  of the tensor network \cite{arad_quantum_2010}. The graphical calculus becomes valuable when using circuit identities that simplify the tensor network. Specifically, these identities encode two simple facts illustrated on \fig{tn_simp}: a permutation $G$ acting on the uniform distribution returns the uniform distribution $ G e^{\otimes t} = e^{\otimes t}$, and a permutation acting on a basis vector returns another basis vector $ Gx_1^N = y_1^N$. 
      
            Once these circuit identities are applied to the evaluation of \eq{Wi} in the specific case of polar codes, it was shown in \cite{ferris_branching_2014,ferris_convolutional_2017} that the resulting tensor network is a tree, so it can be efficiently evaluated. Convolutional polar codes were introduced based on the observation that \eq{Wi} produces a tensor network of constant tree-width despite not being a tree (see \fig{causal_width}), an observation first made in the context of quantum many-body physics \cite{evenbly_class_2014}, so they can also be decoded efficiently. 
      
    \section{Polar code generalizations}
    \label{sec:generalizations}

        In this section, two possible generalizations of polar codes are described and their decoding complexity is analyzed.

        \subsection{Breadth}
        \label{sec:kernels}

            Channel polarization can be achieved using various kernels. In fact, as long as a kernel is not a permutation matrix on $\mathbb F_2^b$, it achieves a non-trivial polarization transform \cite{korada_polar_2010-1}. The CNOT gate is one such example that acts on two bits. However, a general kernel of breadth $b$ can act on $b$ bits, (see \fig{subfig_circ_3_1} for an illustration with $b=3$). An increasing breadth can produce faster polarization, i.e. a decoding error probability which decreases faster with the number of polarization steps. 
     
            Indeed, in the asymptotic regime, Arikan \cite{arikan_channel_2009} showed  that provided the code rate is below the symmetric channel capacity and that the location of the frozen bits are chosen optimally, the asymptotic decoding error probability of the polar code under successive cancellation decoding is $\mathbb P_e \in \mathcal O\qty(2^{N^{-1/2}})$. A different error scaling exponent $\mathbb P_e \in \mathcal O\qty(2^{N^{-\beta}})$ can be achieved from a broader kernel, but breadth 16 is required to asymptotically surpass $\beta=\frac 12$ \cite{korada_polar_2010-1}. 
     
            Such a broad polarization kernel has the drawback of a substantially increased decoding complexity. Arikan \cite{arikan_channel_2009} showed that the decoding complexity of polar codes is $\mathcal O\qty(N \log_2 N)$. From a tensor network perspective, this complexity can be understood \cite{ferris_convolutional_2017} by counting the number of elementary contractions required to evaluate \eq{Wi} and by noting that the tensor network corresponding to \eq{Wi} for $u_i$ and for $u_{i+1}$ differ only on a fraction $1/\log_2 N$ of locations, so most intermediate calculations can be recycled and incur no additional complexity.
            
            As discussed previously, a breadth-$b$ polarization kernel can also be represented as a $2^b\times 2^b$ permutation matrix that act on $\mathbb R^{2^b}$. Applying such a matrix to a $b$-bit probability distribution has complexity $2^b$, and this dominates the complexity of each elementary tensor operation of the successive cancellation decoding algorithm. On the other hand, the total number of bits after $l$ polarization steps with a breadth-$b$ polarization kernel is $N = b^l$, so the overall decoding complexity in this setting is $\mathcal O\qty(2^b N \log_b N)$. 
     
        \subsection{Depth}
        \label{sec:conv}
           
            The previous section described a natural generalization of polar codes which use a broader polarization kernel.  A further generalization, first explored in \cite{ferris_branching_2014,ferris_convolutional_2017}, is to use a polarization step whose circuit is composed of $b$-local gates and has depth $d>1$ (see \fig{subfig_circ_2_2}), which results in a convolutional transformation. 
            A $\cpc_{b,d}$ code, that is, a convolutional polar code with kernel breadth $b$ and circuit depth $d$, is define similarly to a polar code with a kernel of size $b$ where each polarization step is replace by a stack of $d$ polarization layers each shifted relative to the previous layer. 
            \fig{subfig_circ_2_2} and \fig{subfig_circ_3_4} illustrates two realizations of convolutional polar codes.
            
\begin{figure}[!t]
                \centering
                \includegraphics[width=3.5in]{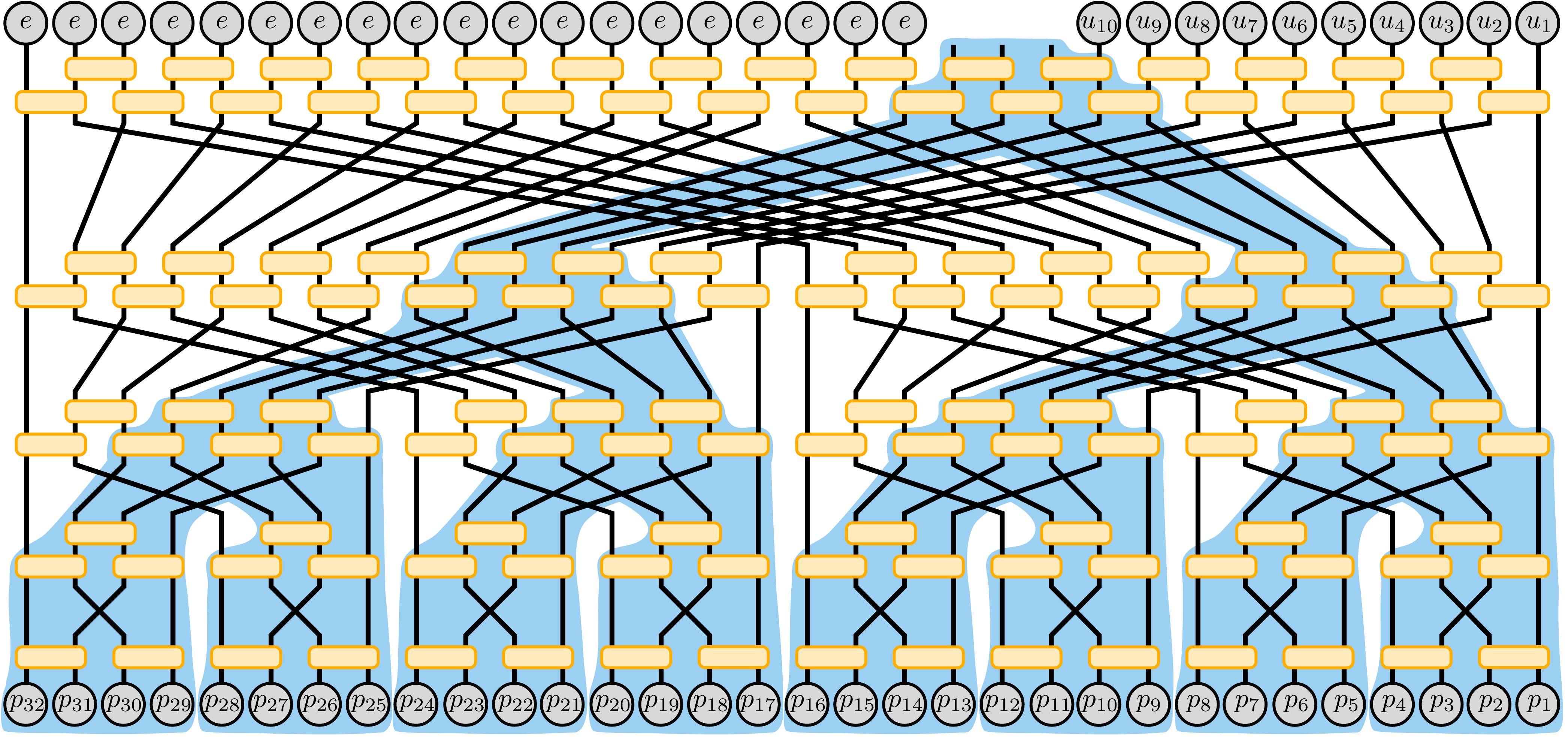}
                \caption{Graphical representation of the causal cone of $u_{11}^{13}$ in the $\cpc_{2,2}$ code. Only the gates in the shaded region receive inputs that depend on the sequence $u_{11}^{13}$. Similarly, the edges contained in the shaded region represent bits at intermediate polarization steps whose value depends on sequence $u_{11}^{13}$. This shows that decoding a $\cpc_{2,2}$ code amouts to contracting a constant tree-width graph. The optimal width $w^* = 3$ and at most $m(2,2,w^*) = 5$ gates are involved per polarization step.} 
                \label{fig:causal_width} 
            \end{figure}

            To analyze the decoding complexity, it is useful to introduce the concept of a causal cone. Given a circuit and a $w$-bit input sequence $u_i^{i+w-1}$, the associated causal cone is defined as the set of gates together with the set of edges of this circuit whose bit value depends on the value of $u_i^{i+w-1}$. Figure \ref{fig:causal_width} illustrates the causal cone of the sequence $u_{11}^{13}$ for the code $\cpc_{2,2}$. 

            Given a convolutional code's breadth $b$ and depth $d$, define $m(d,b,w)$ to be the maximum number of gates in the causal cone of any $w$-bit input sequence of a single polarization step. Because a single convolutional step counts $d$ layers, define $m_s(d,b,w)$ as the number of those gates in the causal cone which are in the $s$-th layer (counting from the top) of the convolution. For the first layer, have $m_1(d,b,w) = \lceil\frac{w-1}b\rceil +1$. This number can at most increase by one for each layer, i.e., $m_{s+1}(d,b,w) \leq m_{s}(d,b,w)+1$, leading to a total number of gates in the  causal cone of a single polarization step 
            \begin{align}
                m(d,b,w) &= \sum_{s=1}^d m_s(d,b,w) \leq dm_1(d,b,w)+\frac{d(d-1)}2 \nonumber \\
                &= d\left\lceil\frac{w-1}b\right\rceil +\frac{d(d+1)}2.
                \label{eq:max_gates}
            \end{align}

            Similarly,  define the optimal decoding width $w^*(b,d)$ as the smallest value of $w$ for which the causal cone of any $w$ bit sequence after one step of polarization contains at most $bw$ output bits.  Figure \ref{fig:causal_width} illustrates that $w^* = 3$ for a $\cpc_{2,2}$ code since any 3 consecutive input bits affect at most 6 consecutive bits after one polarization step. Choosing a decoding width $w^*(b,d)$ thus leads to a recursive decoding procedure which is identical at all polarization steps.  Since the bottom layer counts $m_d(d,b,w) \leq \lceil\frac{w-1}b\rceil +d$ gates, each acting on $b$ bits, we see that there are at most $b\lceil\frac{w-1}b\rceil +db\leq w+db$ output bits in the causal cone of a single polarization step. The optimal decoding width $w^*$ is chosen such that this number does not exceed $bw^*$, thus 
            \begin{equation}
                w^*(b,d) \leq \frac b{b-1}d.
                \label{eq:max_w}
            \end{equation}
            
            Using this optimal value in \eq{max_gates} bounds the number of rank-$b$ tensors that are contracted at each polarization layer, and each contraction has complexity $2^b$. Here again, only a fraction $1/\log_b N$ of these contractions differ at each step of successive cancellation decoding, leading to an overall decoding complexity 
            \begin{align}
                C_{b,d}(N) = 2^b \frac{m(b,d, w^*)}{w^*} N \log_b N\in \mathcal O(2^b d N \log_b N). \label{eq:complexity}
            \end{align}
            Ref. \cite{ferris_convolutional_2017} provides analytical arguments that the resulting convolutional polar codes have a larger asymptotic error exponent $\beta>\frac 12$, and present numerical results showing clear performance improvement over standard polar codes at finite code lengths. These advantages come at the cost of a small  constant increased decoding complexity

    \section{Simulation results}
    \label{sec:results}
        Numerical simulations were performed to analyze the performance of codes breadth and depth up to 4. The breadth-2 kernel used was the CNOT, while the breadth-3 and breadth-4 kernels were 
        \begin{align*}
            &G_3 = \begin{pmatrix}
                1 & 0 & 0 \\
                1 & 1 & 0 \\
                0 & 1 & 1 \\
            \end{pmatrix},
            &&G_4 = \begin{pmatrix}
                1 & 0 & 0 & 0 \\
                1 & 1 & 0 & 0 \\
                0 & 1 & 1 & 0 \\
                0 & 0 & 1 & 1
            \end{pmatrix},
        \end{align*}
        where these are given as representations over $\mathbb F_2^b$. It can easily be verified that these transformations are not permutations, so they can in principle be used to polarize \cite{korada_polar_2010-1}. Also, we choose a convolutional structure where each layer of gates is identical but shifted by one bit to the right (from top to bottom), c.f. \fig{subfig_circ_3_4}. Many others kernel and convolutional structures have been simulated, but those gave the best empirical results.
        
        The encoding circuit $G$ is used to define the code, but the complete  definition of a polar code must also specify the set of frozen bits $\mathcal F$, i.e. the set of bits that are fixed to $u_i=0$ at the input of the encoding circuit $G$. In general, for a given encoding circuit $G$, the set $\mathcal F$ will depend on the channel and is chosen to minimize the error probability under successive cancellation decoding. Here, a simplified channel selection procedure which uses an error detection criteria described in the next section was used. All the simulations presented focus on the binary symmetric memoryless channel.

        \subsection{Error detection}
        \label{sec:ED}

            Considering an error detection setting enables an important simplification in which the channel selection and code simulation can be performed simultaneously without sampling. In this setting, it is consider that a transmission error $x_1^N \rightarrow y_1^N = x_1^N + \vb e$ is not detected if there exists a non-frozen bit $u_i$, $i\in \mathcal F^c$ which is flipped while none of the frozen bits to its right $u_j$, $j<i$, $j\in \mathcal F$ have been flipped. In other words, an error is considered not detected if its first error location (starting from the right) occurs on a non-frozen bit. Note that this does not correspond to the usual definition of an undetectable error which would be conventionally defined as an error which affects no frozen locations. By considering only frozen bits to the right of a given location, the notion used is tailored to the context of a sequential decoder. Empirically, it was observed that this simple notion is a good proxy to compare the performance of different coding schemes under more common settings.

            Denote $\mathbb P_U(i)$ the probability that the symbol $u_i$ is the first such undetected error. Then, given a frozen bit set $\mathcal F$, the probability of an undetected error is $\mathbb P_U = \sum_{i\in \mathcal F^c} \mathbb P_U(i)$. This can be evaluated efficiently using the representation of the encoding matrix over $\mathbb R^{2^N}$ as above. For $\vb e \in \mathbb F_2^N$, denote $\mathbb P(\vb e)$ the probability of a bit-flip pattern $\vb e$, viewed as a vector on $\mathbb R^{2^N}$. At the output of the symmetric channels with error probability $p$, $\mathbb P^T = (1-p,p)^{\otimes N}$. Then
            \begin{equation}
                \mathbb P_U(i) = (1-p,p)^{\otimes N} G \binom 10^{\otimes i-1} \otimes \binom 01\otimes e^{\otimes(i-1)},
                \label{eq:detection}
            \end{equation}
            where here again $e = \binom 11$. In terms of tensor networks, this corresponds to the evaluation of the network of \fig{subfig_eff_channel} with $u_i = 1$ and all $u_j=0$ for all $j<i$. Thus, this can be accomplished with complexity given by \eq{complexity}.

            Because the evaluation of \eq{detection} is independent of the set of frozen bits, it can be evaluate for all positions $i$, selecting the frozen locations as the $N-K$ locations $i$ with the largest value of $\mathbb P_U(i)$. Then, the total undetected error probability is the sum of the $\mathbb P_U(i)$ over the remaining locations. This is equivalently the sum of the $K$ smallest values of $\mathbb P_U(i)$.

            \begin{figure*}[!th]
                \centering
                \subfloat[]{
                    \centering
                    \includegraphics[width=2.25in]{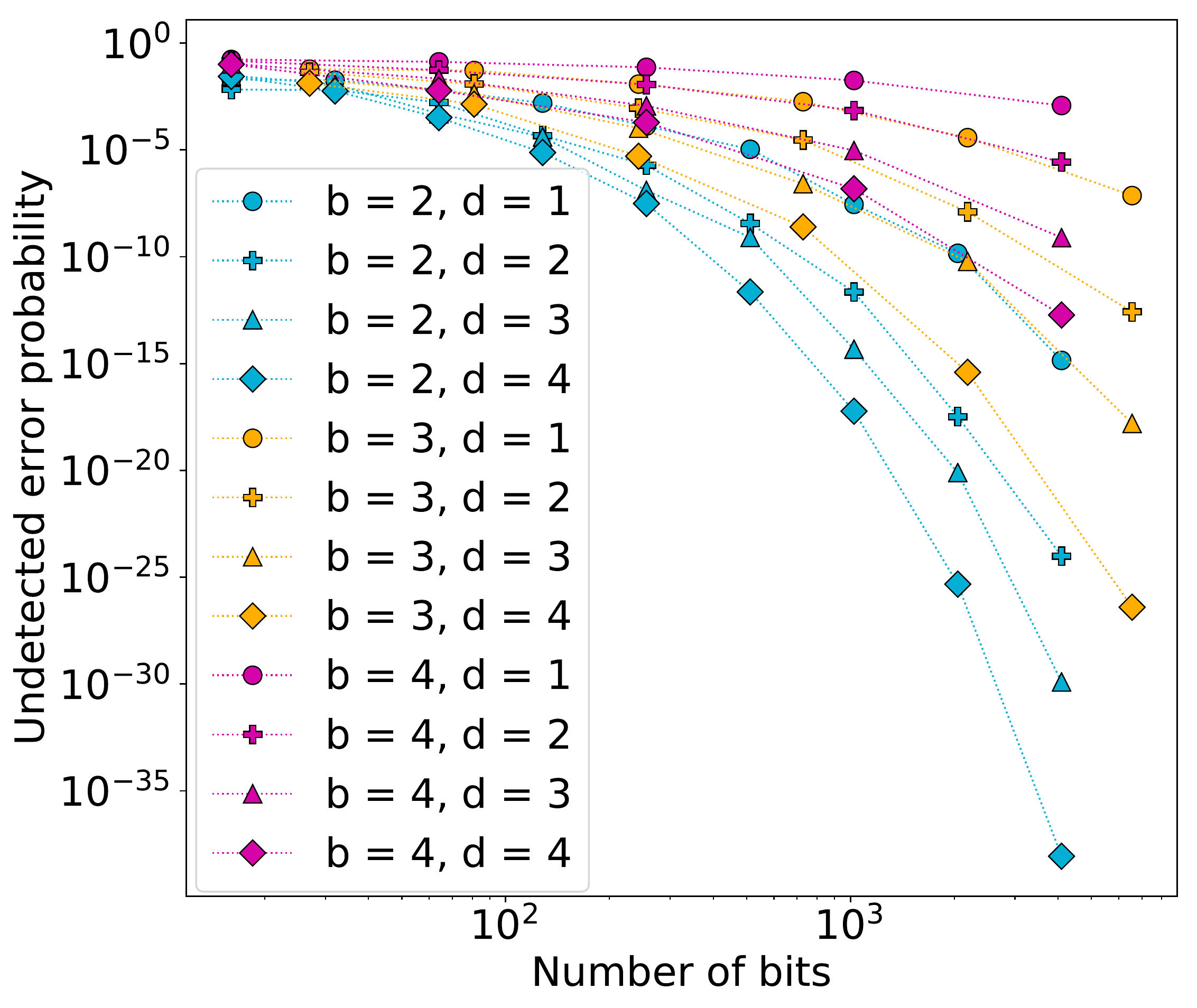}
                    \label{fig:subfig_erasure_all}}
                \subfloat[]{
                    \centering
                    \includegraphics[width=2.25in]{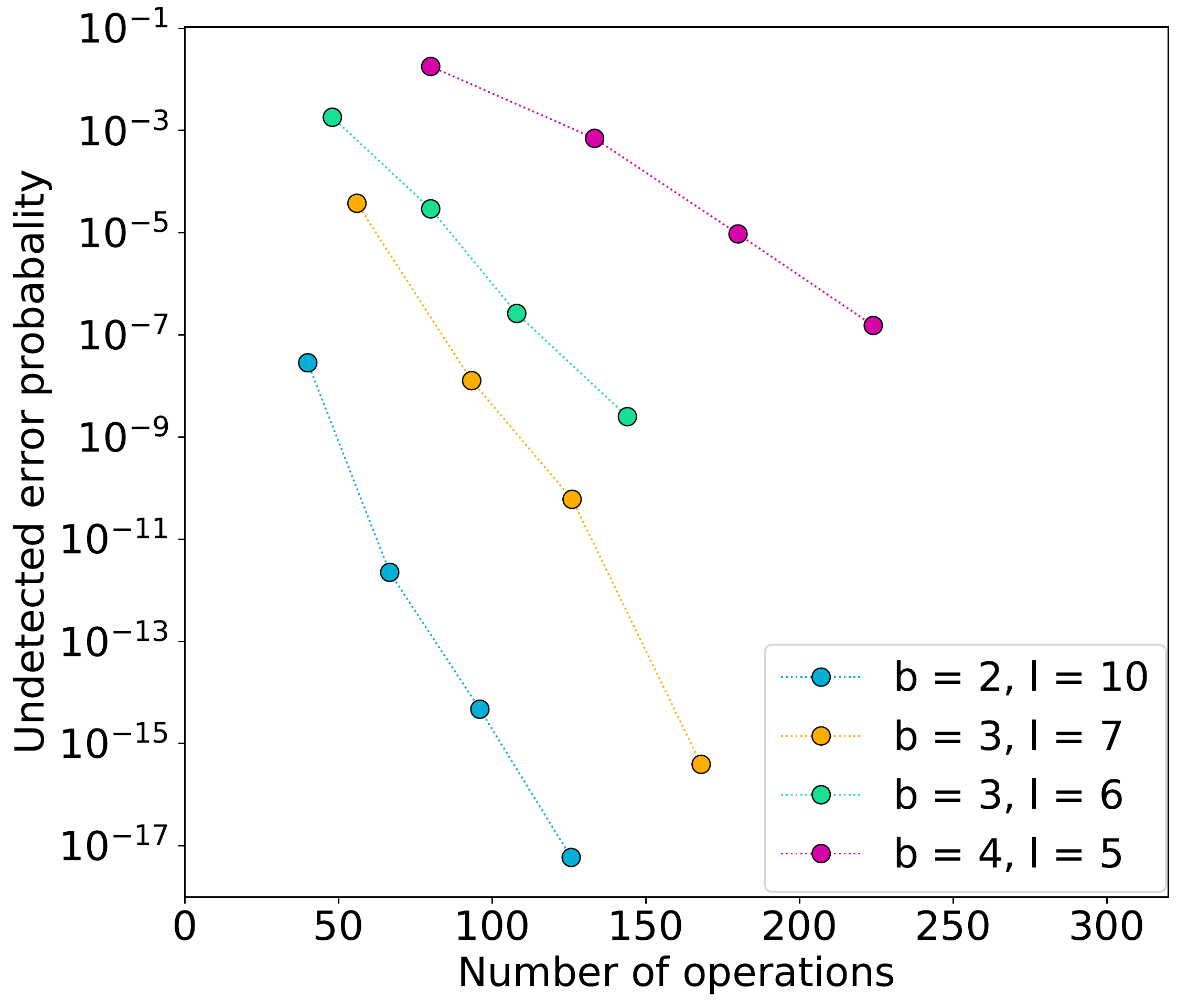}
                    \label{fig:subfig_erasure_complexity}}
                \subfloat[]{
                    \centering
                    \includegraphics[width=2.25in]{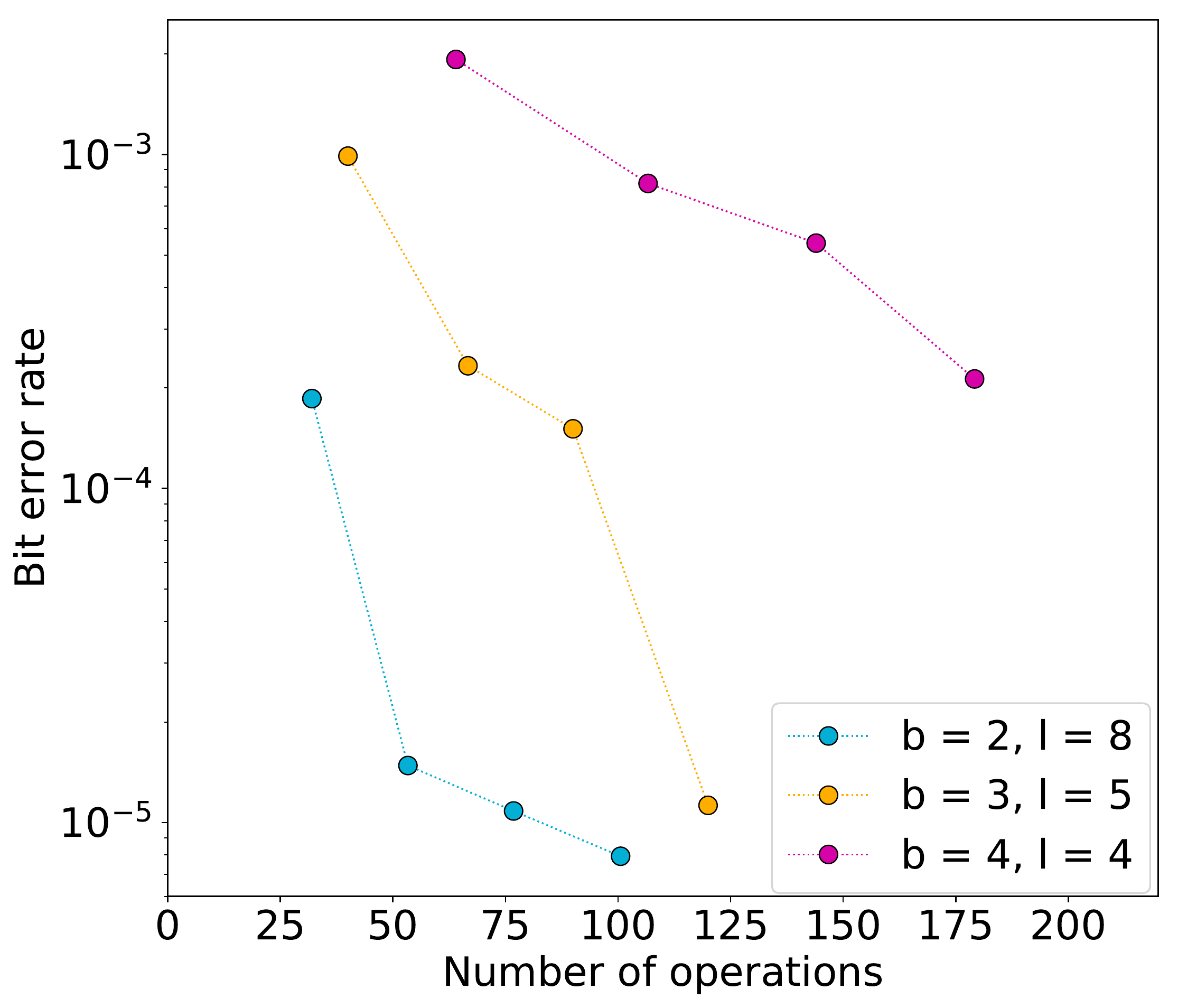}
                    \label{fig:subfig_flip_complexity}}
                \quad
                \caption{Numerical simulation results. \textbf{(a)} Undetected error probability under successive cancellation decoding for polar codes ($d=1$) and convolutional polar codes ($d>1$) for various kernel breadths $b$, plotted as a function of code size $N=b^l$ by varying the number of polarization steps $l$. The channel is BSC($1/4$) and the encoding rate is $1/3$. \textbf{(b)} Same thing as (a) but plotted as a function of their decoding complexity, c.f. \eq{complexity}. The number of polarization steps $l$ is chosen in such a way that all codes are roughly of equal size $N(b,l)=b^l\approx 10^3$. The dots connected by a line all have the same kernel breadth $b$ but show a progression of depth $d=1,2,3,4$, with $d=1$ appears on the left and corresponds to regular polar codes. \textbf{(c)} The bit error rate for a BSC($1/20$) with an $1/3$ encoding rate plotted as a function of the decoding complexity. The depth is specify similarly to (b) by the connected dots. The number of polarization steps is chosen to have roughly $N \approx 250$ bits.}
                \label{fig:results} 
            \end{figure*}

            The results are shown on \fig{subfig_erasure_all} for various combinations of kernel breadths $b$ and convolutional depth $d$. The code rate was $\frac 13$, meaning that the plotted quantity is the sum of the $N/3$ smallest values of  $\mathbb P_U(i)$. \fig{subfig_erasure_complexity} presents a subset of the same data with parameters $b$ and $l$ resulting in codes of roughly equal size $N = b^l\approx 10^3$. This implies that codes with larger breadth use fewer polarization steps. The undetected error probability $\mathbb P_U$ is then plotted against the decoding complexity, compute from \eq{complexity}. Notice that increasing the depth is a very efficient way of suppressing errors with a modest complexity increase. In contrast, increasing the breadth actually deteriorates the performance of these finite-size code and increases the decoding complexity. 

        \subsection{Error correction} 
        \label{sec:EC}

            For the symmetric channel, the frozen bits were chosen using the error detection procedure describe in the previous section. This is not optimal, but it is sufficient for the sake of comparing different code constructions. Then, standard Monte Carlo simulations were done by transmitting the all 0 codeword sampling errors, using successive cancellation decoding and comparing the decoded message. The results are presented in \fig{subfig_flip_complexity}. The conclusions drawn from the error detection simulations all carry over to this more practically relevant setting.

    \section{Conclusion}
    \label{sec:conclusion}
        
        We numerically explored a  generalization of the polar code family based on  a convoluted polarization kernel given by a finite-depth local circuit. On practically relevant code sizes, it was found that these convoluted kernel offer a very interesting error-suppression {\em vs} decoding complexity trade-off compare to previously proposed polar code generalizations using broad kernels. Empirically, no incentive were found to consider increasing both the breadth and the depth: an increasing depth alone offers a greater noise suppression at comparable complexity increase. It will be interesting to see what further gains can be achieved, for instance, from list decoding of convolutional polar codes. 

    \section*{Acknowledgment}

       This work was supported by Canada's NSERC and Québec's FRQNT. Computations were done using Compute Canada and Calcul Québec clusters.

    \bibliographystyle{IEEEtran}
    \bibliography{IEEEabrv,ref}
    
\end{document}